\documentclass[a4paper,showkeys,showpacs]{revtex4}

\usepackage[english]{babel}
\usepackage{amssymb}
\usepackage{amsbsy}
\usepackage{amstext}
\usepackage{graphicx}
\usepackage{subfigure}
\usepackage{pst-node}
\usepackage{verbatim}
\usepackage{ulem}   

\newcommand{\be}{\begin{eqnarray}}
\newcommand{\ee}{\end{eqnarray}}
\newcommand{\bdm}{\begin{displaymath}}
\newcommand{\edm}{\end{displaymath}}
\newcommand{\ds}{\displaystyle}
\newcommand{\ba}{\begin{array}}
\newcommand{\ea}{\end{array}}
\newcommand{\pa}[1]{\left(#1\right)}

\newcommand{\omm}{{\omega_m}}

\newcommand{\omrd}{{\omega_{rd}}}

\begin{document}

\title{Phenomenological gravitational waveforms from spinning coalescing 
  binaries}

\author{R. Sturani$^{(1,2)}$, S. Fischetti$^{(3)}$, L. Cadonati$^{(3)}$, G. M. 
Guidi$^{(1,2)}$, J. Healy$^{(4)}$, D. Shoemaker$^{(4)}$, A. Vicer\'e$^{(1,2)}$}

\affiliation{(1) Dipartimento di Scienze di Base e Fondamenti, Universit\`a 
  degli Studi di Urbino `Carlo Bo', I-61029 Urbino, Italy\\
  (2) INFN Sezione di Firenze, I-50019 Sesto Fiorentino, Italy\\
  (3) Physics Department, University of Massachusetts, Amherst MA 01003, USA\\
  (4) Center for Relativistic Astrophysics, Georgia Tech, Atlanta, GA 30332, USA}

\email{riccardo.sturani@uniurb.it}

\begin{abstract}
  An accurate knowledge of the coalescing binary gravitational waveform
  is crucial for experimental searches as the ones 
  performed by the LIGO-Virgo collaboration.
  Following an earlier paper by the same authors we refine the construction
  of analytical phenomenological waveforms describing the signal sourced 
  by generically spinning binary systems.
  The gap between the initial inspiral part of the waveform, described by 
  spin-Taylor approximants, and its final ring-down part, described by damped 
  exponentials, is bridged  by a phenomenological phase calibrated by 
  comparison with the dominant spherical harmonic mode of a set of waveforms 
  including both numerical and phenomenological waveforms of different type.
  All waveforms considered describe equal mass 
  systems.
  The Advanced LIGO noise-weighted overlap integral between the numerical and 
  phenomenological waveforms presented here ranges between 0.95 and 0.99 for a 
  wide span of mass values.
\end{abstract}


\maketitle

\section{Introduction}

The experimental program for gravitational wave detection is on the way, as the 
network of kilometer-scale interferometers formed by the Laser 
Interferometer Gravitational-wave Observatory (LIGO), Virgo and GEO 
are presently in science run or undergoing substantial upgrades taking them 
to their advanced version \cite{Abbott:2007kv,Acernese:2008zzf,Willke:2007zz}.

Among the possible sources one of the most promising is represented by 
the coalescence of compact binary systems of neutron stars or black holes.
The coalescence of binary systems is usually described in terms of three
distinct phases: the {\it inspiral}, the {\it merger} and the 
{\it ring-down}. The inspiral phase allows for an accurate analytical 
description via the so-called Post-Newtonian (PN) expansion, see for 
instance \cite{Blanchet:2006zz} for a review. 
The ring-down also admits a perturbative analytical model, as it describes the 
damped oscillations of the single object resulting from the binary coalescence,
as a superposition of black-hole quasi-normal modes \cite{Teukolsky:1973ha}. 
The merger phase is however fully non-perturbative and for generic systems it 
has not been described analytically but rather by numerical simulations. 

During the last six years numerical relativity has made tremendous 
progress in describing the full coalescence of a binary system 
beginning with \cite{Pretorius:2005gq,Campanelli:2005dd, Baker:2005vv} and more
recently~\cite{Herrmann:2007ex,Tichy:2008du,Scheel:2008rj,Baker:2008mj,Ajith:2007qp,Lousto:2009ka,Pollney:2007ss}, 
see~\cite{Hannam:2009rd,Husa:2007zz,Pretorius:2007nq} for reviews,
and it can now produce waveforms for generic spin orientations, with moderate 
spin magnitude ($\lesssim 0.9$) and mass ratios ($\lesssim 10:1$).

Match-filtering techniques represent a powerful tool for seeking signals,
but they need a detailed knowledge of the waveform in order to have high
efficiency and to be useful for parameter estimation. 
They are extensively used in LIGO-Virgo experimental searches
in order to uncover weak signals buried into noise, see e.g. 
\cite{Abadie:2011kd} for the results of a recent match-filtered search.
In such searches real data are compared with banks of template waveforms, made
of a large number (tens of thousands) of templates: due to the computational 
cost of numerical simulations, it would be impractical to numerically generate
the waveforms necessary to populate template banks.

There exist nowadays analytically constructed waveforms describing the 
entire coalescence of a binary system. They have been achieved 
in the Effective One Body (EOB) construction 
\cite{Buonanno:1998gg,Buonanno:2006ui,Buonanno:2007pf,Damour:2009kr}
for non-spinning systems, and in the 
EOB-spin waveforms \cite{Pan:2009wj}, for binaries with spins aligned with the 
orbital angular momentum (thus non-precessing spins). In both cases of spinning 
and non spinning waveforms, a comparison with numerically generated waveforms 
is needed by the EOB construction in order to calibrate some free parameters of
the model.
Another method for generating analytical waveforms for
spinning non-precessing binaries is by joining PN-generated inspiral with a fit
of numerical waveforms  to construct {\it{phenomenological}} waveforms, as done
in \cite{Ajith:2007qp}.
In this paper, we further investigate on the family of phenomenological 
waveforms introduced in \cite{Sturani:2010yv}, dubbed PhenSpin,  
which are analytical waveforms describing the entire coalescence of 
{\it generically spinning} compact binaries.
In particular a restricted set of {\it phenomenological} waveforms 
\cite{Ajith:2007qp} describing non-precessing systems are used here
together with the same set of fully precessing numerical waveforms used in 
\cite{Sturani:2010yv} to tune this new version of PhenSpin waveforms.

The PhenSpin waveforms have been constructed by 
joining the perturbative PN description of the inspiral to the ring-down phase 
by a phenomenological phase which plausibly describes the evolution of the 
waveform in between. 
With respect to the previous work introducing these waveforms, here we 
give a slightly modified (improved) version, identical in spirit but
better tuned in some technical aspects, and produce new results assessing their
faithfulness to numerical simulation in presence of detector noise.
The improved details allow to obtain a slightly better match between
the PhenSpin waveforms and the set of test waveforms they have been compared 
with.

Waveforms describing generically spinning coalescing binaries are not suitable
for searches employing match-filtering, as 
the size of a template bank increases exponentially with the number of template
parameters: since spinning waveforms depend on several parameters (masses, 
spin components of the binary constituents, angles defining the orientation
of the source with respect to the observer) it is not practical 
to construct one template bank to cover the entire spin parameter space. 
Non spinning, or at least non-precessing waveforms, are usually preferred for 
template bank construction, with the exception of the so-called Physical 
Template Family, representing a single spin family waveform \cite{Pan:2003qt}, 
which however can effectively describe also doubly spinning physical systems 
\cite{Buonanno:2004yd}.

The availability of generically spinning waveforms is however badly needed
to assess the efficiency of experimental searches based on banks of 
non-spinning templates, like \cite{Abadie:2011kd}.
Moreover fully spinning waveforms can be used as templates in connection
with parameter estimation via Bayesian inference methods, which can be used as 
follow-up analysis to perform searches in the parameter space with full 
dimensionality, but restricted to a small subset of the entire space, as 
determined by lower level triggers.

The paper is organized similarly to \cite{Sturani:2010yv} and the
exposition has been kept here as self-contained as possible.
In sec.~\ref{se:method} the analytical waveform construction and 
the waveforms used for calibration are revisited. Differences with respect
to the old version of PhenSpin are described.
In sec.~\ref{se:result} the results are presented, in the form of comparison 
between analytically and numerically generated waveforms, which reproduce the 
dominant quadrupolar mode $l=m=2$.
In sec.~\ref{se:concl} the conclusions that can be drawn from the present 
work are reported.

\section{The method}
\label{se:method}

Following the original introduction of PhenSpin waveforms given in 
\cite{Sturani:2010yv}, 
the present work revisits the construction of analytical gravitational 
waveforms generated by the coalescence of spinning binary systems. 
The waveforms used to construct and calibrate our analytical model
include the numerical waveforms used in the previous paper,
describing equal mass binary systems ($m\equiv m_1=m_2$), with spin 
magnitudes $|{\mathbf {S_1}}|=|\mathbf{S_2}|=0.6\, m^2$ and starting with
$\mathbf{S_2}$ 
orthogonal to the initial orbital angular momentum (where $\mathbf{S_{1,2}}$
denote the binary constituent spin vectors and we posit $G_N=c=1$).
Here in addition, to cover a portion of the parameter space not addressed by 
the numerical simulations, we use also four
phenomenological waveforms of the type described in 
\cite{Ajith:2007qp}, generated via the LAL libraries
\cite{lal} to calibrate the PhenSpin in the case of
aligned spins, which together with the numerical relativity waveforms
form our set of {\it test waveforms}. 

The description of the dynamics adopted here models the inspiral phase via 
the standard TaylorT4 PN formulae, see \cite{Buonanno:2009zt} for definition 
and comparison of different PN approximants in the spin-less case. 
Un-like the non-precessing case, the knowledge of the time-varying amplitude 
and phase is necessary but not sufficient to determine the waveform from
spinning precessing binaries, as it must be complemented by the 
spin and angular momentum evolution, see e.g. \cite{Arun:2008kb}:
\renewcommand{\arraystretch}{1.4}
\be
\ba{rcl}
\ds{\mathbf{\dot S_{1,2}}}&=&
\ds{\mathbf \Omega_{1,2}}\times {\mathbf S_{1,2}}\,,\\
\ds{\mathbf{\dot{\hat L}}}&=&
\ds-\frac \nu v \pa{\mathbf{\dot S_1}+{\mathbf{\dot S_2}}}\,,
\ea
\ee
\renewcommand{\arraystretch}{1.}
where  
\be
\mathbf \Omega_{1,2}\equiv \pa{\frac 34 +\frac \nu 2\mp \frac 34\delta}
{\mathbf{\hat L_N}}
\ee
and $\nu\equiv m_1m_2/M^2$ is the symmetric mass ratio and 
$\delta\equiv (m_1-m_2)/M$, being $M\equiv m_1+m_2$ the total mass of the 
binary system and ${mathbf \hat L}$ the orbital angular momentum unit vector.
It is convenient to define an {\it orbital} phase 
$\phi=\int \omega_{orb}\,dt$ whose evolution is given by
\begin{equation}
v^3 \equiv \omega_{orb} M\,, \qquad  \frac{dv}{dt}=-\frac{F(v)}{dE/dv}\,,
\end{equation}
where $F(v)$ and 
$E(v)$ are respectively the flux emitted and the energy of a circular orbit
with angular frequency $\omega_{orb}$, related to the main gravitational wave 
frequency $f_{GW}$ via $f_{GW}=\omega_{orb}/\pi$.\\
By parametrizing the orbital angular momentum unit vector $\hat{\mathbf L}$ as 
\be
{\mathbf{\hat L}=(\sin\iota\cos\alpha,\sin\iota\sin\alpha,\cos\iota)}
\ee
it is convenient to introduce the {\it carrier} phase $\Psi$ given by
\be
\frac {d\Psi}{dt}=\omega_{orb}-\cos\iota\,\frac{d\alpha}{dt}\,.
\ee

Numerically generated waveforms are usually decomposed in spherical harmonics,
in particular the five quadrupolar modes ($l=2$) are the only non-vanishing
at the lowest order in $v$, and the $l=2,m=\pm 2$ mode are the dominant ones. 
As determined by the PN analysis, the $l=2,m=2$ mode 
in the inspiral phase $h^{(insp)}_{2,2}$ (the only one which
will be used here for comparison with test waveforms, which can be expressed in 
terms of the usual plus and cross polarizations as 
$h^{(insp)}_{2,2}=h_+-ih_\times$)
is given by formulae which can be found e.g. in \cite{Arun:2008kb}, which 
here we re-write in the following form
\renewcommand{\arraystretch}{1.4}
\be
\label{eq:h22}
\ba{rl}
\ds h_{2,2}^{(insp)}(t) = -2\sqrt{\frac{16\pi}{5}}\frac{\nu M}d 
v^2 & \left[\ds \pa{1-c(v)\frac{v^2}{42}(107-55 \nu)}
\left(\cos^4\pa{\iota/2}e^{-2i(\Psi+\alpha)}+
\sin^4(\iota/2)e^{2i(\Psi-\alpha)}\right)+\right.\\ 
& \ds \left. v \frac \delta 3 \sin\iota 
\pa{\sin^2\frac\iota 2e^{i(\Psi-2\alpha)} + \cos^2\frac\iota 2
e^{i(\Psi+2\alpha)} } \right]+O(v^5)\,,
\ea
\ee
\renewcommand{\arraystretch}{1}
where spin-dependent terms in the amplitude have been 
neglected as well as terms of order higher than $v^4$, $t$-dependence is 
understood in $\Psi$ and $\alpha$ and the $z-$axis used for defining $l,m$ 
modes is parallel to the initial total angular momentum. 

Note that differently from \cite{Arun:2008kb} and the previous version of the 
PhenSpin waveforms, the $v^4$ terms have been added weighting them
by a phenomenological function $c(v)$ defined as
\bdm
c(v)= \left\{
\ba{cl}
\exp\left[-(1-0.05/\omega_{orb})^2/2\right]&\omega_{orb}\leq 0.05\\
1&\omega_{orb}>0.05\\
\ea
\right.
\edm
which has the role of turning $v^4$ corrections on at values of the orbital 
frequency $M\omega_{orb}\gtrsim 0.05$, as otherwise poor matching with 
numerical simulations would be obtained.

The $m=-2$ mode can be obtained via
$h_{2,-2}(\Psi)=h^*_{2,2}(\Psi+\pi)$ and in the equal mass case
$h_{2,-2}=h^*_{2,2}$ holds: we thus focus the calibration of our phenomenological
model on the $h_{2,2}$ mode.\\
The functions $F(v)$ and $E(v)$ are necessary to determine the orbital phase and
they are 
known up to 3.5PN order as far as orbital effects are concerned, and up to 
3PN and 2PN level for respectively ${\mathbf{S_{1,2}}}{\mathbf L}$ and
$\mathbf{S_1}\mathbf{S_2}$, $\mathbf{S_1}\mathbf{S_1}$,
$\mathbf{S_2}\mathbf{S_2}$ interactions, see 
\cite{Mikoczi:2005dn,Racine:2008kj,Blanchet:2011zv}
for recent derivations of spin-orbit and spin-spin interaction effects.

According to studies in the non-spinning case 
\cite{Buonanno:2006ui,Baker:2006ha,Boyle:2007ft},
the TaylorT4 appears to be a very good approximant up to a frequency 
$\bar \omega = \pi\bar f_{GW}\simeq 0.1/M$ for the equal mass case,
even though its faithfulness seems to worsen for different mass-ratios (which
however are not considered in this work). 
\footnote{For comparison $f_{GW}\simeq 65 Hz \pa{\frac{\omega M}{0.01}}
\pa{\frac M{10M_\odot}}^{-1}$.} 

The PN evolution (\ref{eq:h22}) is halted at $t=t_m$, when $\omega_{orb}$ 
reaches the value
$\omm$ that is determined by comparison with the test waveforms. 
For $\omega_{orb}>\omm$
($\omega_{orb}$ is monotonically increasing) the angular frequency is evolved 
according to 
\be
\label{eq:omfit}
\omega_{orb}(t)=\frac{\omega_1}{1-t/T_A}+\omega_0,\quad 
\omm<\omega_{orb}< z\,\omrd\ {\rm and}\ t_m<t<t_{rac}\,,
\ee
where the three unknown parameters $\omega_{0,1}$ and $T_A$ are determined by 
requiring continuity of $\omega_{orb}$ and its first and 
second derivatives at the matching point defined by $\omm$.\\
The damped exponentials describing the ring-down phase are attached at the 
instant of time $t_{rac}$ when $\omega_{orb}$ reaches a fraction $z$ 
of the $\omrd$ value, the specific value of $z$ has been determined like
$\omm$ by comparison with the numerical relativity waveforms, as described
below, and differently with respect to the previous version of the PhenSpin 
waveforms, where $z$ has not been fit to numerical simulations but kept
constant to a convenient value.

Differently from the first PhenSpin version \cite{Sturani:2010yv}, where it 
has been kept constant at their value at $t=t_m$, the angular variable 
$\alpha$ is evolved with a similar phenomenological formula 
\renewcommand{\arraystretch}{1.5}
\be
\displaystyle\frac{d\alpha}{dt}&=&\displaystyle
\frac{\dot\alpha_1}{1-t/T_A}+\dot\alpha_0\,,
\ee
\renewcommand{\arraystretch}{1}
where the parameters $\dot\alpha_{1,0}$ are determined by 
requiring the continuity of $\alpha$ up to its second derivative,
and $T_A$ is the same as determined in eq.(\ref{eq:omfit}).

Finally the usual ring-down description of the waveform is used
\be
\label{ringdown}
\ba{rcl}
h^{(rd)}_{2,2}(t)&=&\ds \sum_ne^{-t/\tau_n}A_ne^{i{\omrd}_n t}\quad t>t_{rac}\,,\\
\ea
\ee
where it is used that the ring-down phase is described by adding damped 
exponentials of increasing inverse damping time, the overtones, with complex 
constants $A_n$'s. Here we assume that given the moderate spin values we are 
considering, the direction of the final spin of the black hole is parallel to 
the initial total angular momentum. 
We have checked that during 
the PN-inspiral phase ($t<t_m$) this is indeed the case to very accurate 
precision (better than $10^{-4}$) for all spin configurations considered 
here. In \cite{Fischetti:2010hx} the same numerical simulations used here are 
analyzed and a maximum misalignment angle $\theta$ between the final spin and 
the initial total angular momentum is found to be around 
$\theta\simeq 0.24 {\rm rad}\simeq 13^o$
(see fig.3 of \cite{Fischetti:2010hx}).
Further investigations are necessary to assess the importance of this effect on
the actual waveform shape.

Allowing more overtones requires to fix more coefficients, 
which can be done by admitting continuity of the waveform to the appropriate 
level: using $n$ overtones requires matching the waveform up to its 
$2\times(n-1)$-th derivative, as each overtone involve the determination of
a complex (or two real) constant parameter(s).
The construction that inspired our work is the EOB matched to numerical
relativity waveforms (usually referred to as EOBNR), introduced in 
\cite{Buonanno:2006ui} where the  waveform is assumed 
circularly polarized (i.e. $h_\times(\Psi+\pi/4)=\pm h_+(\Psi)$), 
so that the real and imaginary part of $A_n$ for the $l=2,m=2$ are not 
independent parameters. Here however 
we do not assume circular polarization, as in general terms of order $v^3$ in 
eq.(\ref{eq:h22}) (for unequal mass systems) will spoil this property and the 
stitching of the ring-down modes is performed independently on the real and
imaginary part of each multipolar mode.
For any such mode defined by a $(l,m)$ pair there is an infinity of 
overtones with increasing damping factors, but for our practical purposes 
retaining only two overtones is enough.

As described in \cite{Leaver:1985ax} each overtone with given $l,m$ will be 
in general a superposition of the two modes which are usually designated by
$l,m$ and $l,-m$. Here we stick to the prescription adopted in 
\cite{Buonanno:2006ui} where only the $m>0$ mode is stitched to the inspiral
waveform. We have verified that adding the $l=2,m=-2$ mode to $h^{(rd)}_{2,2}$ 
would not improve the fit to the numerical waveforms, at the expense of 
introducing additional constant parameters which have to be fixed by imposing 
further continuity requirements.

The values of the ring-down frequencies and damping factors of
the three lowest overtones of the $l\leq 4$ modes can be read from 
\cite{Berti:2005ys} as a function of the mass and spin of the final object 
created by the merger of the binary system.
We estimate the final mass by taking the algebraic sum of the constituents' 
masses and the negative binding energy once $\omm$ is reached, and the final 
spin according to the phenomenological formula given in eq.(5) of 
\cite{Barausse:2009uz}.

The analytical waveforms just described have been quantitatively confronted 
with the set of test ones  by computing the noise-less overlap integral 
\be
\label{eq:overlap}
I_{\hat h_1,\hat h_2}\equiv 2 \int_{0}^\infty \pa{\hat h_1(f)\hat h_2^*(f)+
\hat h_1^*(f)\hat h_2(f)}\, df
\ee
maximized over initial phase and time of arrival, 
where normalized waveforms have been considered
\bdm
\hat h(f)\equiv\frac{h(f)} 2\pa{\int_0^\infty |h(f)|^2}^{-1/2}\,.
\edm

The angular frequency $\omm$ and the parameter $z$ have been determined by 
comparison with a 
first set of short test waveforms (4-6 cycles long) by picking 
the values maximizing the overlap integral (\ref{eq:overlap}) (with a precision
respectively of $\pm 5\cdot 10^{-4}/M$ and $0.01$). 
The set of short waveforms have initial orbital frequency 
$\omega_{orb;ini}\sim 0.05/M$ (for comparison, $\omrd\sim 0.3/M$). 
Note that despite $\omega_{orb;ini}$ being not too far from
the values of $\omm$ obtained through the fit, see tab.~\ref{tab:short}, it 
still allows at least one oscillation cycle before the onset of the 
phenomenological phase for all test waveforms.

The numerical waveforms in the set of test waveforms
have been generated with 
\texttt{MayaKranc}.  The grid structure for each run consisted of 10 levels of 
refinement provided by \texttt{CARPET} \cite{Schnetter-etal-03b},
a mesh refinement package for \texttt{CACTUS} \cite{cactus-web}.
Sixth-order spatial finite differencing was used with the BSSN equations 
implemented with \texttt{Kranc}~\cite{Husa:2004ip}.  The outer boundaries are 
located at $317M$ and the finest resolution is $M/77$.  Waveforms were 
extracted at $75M$.
A few waveforms were generated at resolutions of
$\{M/64, M/77, M/90\}$, and convergence consistent with our fourth order
code is found.  The short (long) runs showed a phase error on the order of 
$5\cdot 10^{-3}$ ($5\cdot 10^{-2}$) radians and an amplitude error of 
$\approx 2\%$ ($\approx 5\%$).

The numerical waveforms consist of two 
sets: the first set consisted in 24 few-cycle-long waveforms, representing 
mostly 
the merger and ring-down phases of a coalescence which, together with four
phenomenological ones with aligned spins, has been used as described above to 
fix the values of $\omm$ and $z$ for the corresponding values of initial spins.
All of the numerical waveforms have initial spin 
${\mathbf S_2}/m^2=(-0.6,0,0)$ in the reference frame in which the initial 
$\mathbf{\hat L}=(0,0,1)$. The different values of the first dimension-less 
spin have been obtained by rotating the $(0,0,0.6)$ vector by 15 degrees in the
x-z plane. 
This set of numerical waveforms has been completed by the addition of four 
phenomenological waveforms: one spin-less and three with spins aligned with the
orbital angular momentum ${\mathbf L}$ and the 
same magnitude as above (one with both spins aligned with ${\mathbf L}$, one 
with spins pointing in opposite directions and a third with both 
spins anti-aligned with ${\mathbf L}$).
Once determined the values of $\omm$ and $z$ for each waveform, their values
for generic spins have been determined by
assuming an analytical dependence on the dimension-less spin 
$\mathbf{\chi_{1,2}}$ defined as
${\mathbf \chi_{1,2}}\equiv {\mathbf S_{1,2}}/m_{1,2}^2$, according to
\renewcommand{\arraystretch}{1.4}
\be
\label{eq:omegamatch}
\ba{rl}
\!\!\!\!\!\!\!\!\!\!\!\!\!\! M\omm =& 
a_0 + a_1 (\chi_{1z}+\chi_{2z}) + a_2 \delta (\chi_{1z}-\chi_{2z}) +
a_3 ({\bf \chi_1\chi_2})+\\
&a_4({\bf \chi_1^2}+{\bf \chi_2^2})+a_5 (\chi_{1z}\chi_{2z}) + 
a_6 (\chi_{1z}^2+\chi_{2z}^2) + (\chi_{1z}+\chi_{2z})\times\\
&\left[a_7({\bf \chi_1\chi_2})+
a_8({\bf \chi_1^2}+{\bf \chi_2^2})+a_9(\chi_{1z}\chi_{2z})+ 
a_{10}(\chi_{1z}^2+\chi_{2z}^2)\right]
\ldots\,,
\ea
\ee
\renewcommand{\arraystretch}{1}
where the spin components are understood in a frame where the orbital angular 
momentum is along the $z$ axis and higher powers of the spin components have 
been neglected.
From eq.~(\ref{eq:omegamatch}) one can note that since spins are evolving in 
time, $\omm$ is also slightly changing with time, the explicit values taken by 
the $a_i$ coefficients are reported in tab.~\ref{tab:fit}. 
The $\chi_i$ values used to fit $\omm$ are taken at $t=t_m$.

The dependence of the $\omega_{orb}$ evolution equation on $\mathbf{L}$ and 
$\mathbf{S_{1,2}}$ implies that the spin components parallel to the orbital 
angular momentum enter already at linear level, whereas the dependence on the 
spin components in the plane of the orbit starts from the quadratic level. 
The $a_i$ coefficients may depend on the symmetric mass ratio $\nu$, but it is 
assumed here that they can be analytically expanded around their value at 
$\delta=0$, according to
\be
\label{eq:expansion}
a_i(\delta)=a_i+\delta\, a_i^{(1)}+\delta^2 a_i^{(2)}+\ldots\,.
\ee
Note that anti-symmetric combinations of spin components do not appear for 
$\delta=0$.
Given the specifics of the test waveform set we used (all having $\delta=0$) 
we could not calculate the coefficients $a_i^{(i)}$ nor $a_2$.\\
An analog formula has been assumed for $z$
\renewcommand{\arraystretch}{1.4}
\be
\label{eq:fracmatch}
\ba{rl}
\!\!\!\!\!\!\!\!\!\!\!\!\!\! z =& 
b_0 + b_1 (\chi_{1z}+\chi_{2z}) + b_2 \delta (\chi_{1z}-\chi_{2z}) +
b_3 ({\bf \chi_1\chi_2})+\\
& b_4({\bf \chi_1^2}+{\bf \chi_2^2}) +b_5 (\chi_{1z}\chi_{2z}) + 
b_6 (\chi_{1z}^2+\chi_{2z}^2)\,,
\ea
\ee
\renewcommand{\arraystretch}{1.}
where terms cubic in the spins have not been necessary here.
Results from the fit of $z$ are reported in tab.~\ref{tab:fit}.

These values have then been tested by computing the {\it faithfulness}
of the now 
fully calibrated PhenSpin waveforms with respect to a second set of long
waveforms, consisting of 8 long numerical waveforms (12-15 cycles long) plus 
the 10 phenomenological ones: 4 with the same parameters as above 
(initial $\omega_{orb}\simeq 0.03/M$) and six additional ones
characterized by spins aligned with the angular momentum and chi-pair values
give by $(\chi_{1z},\chi_{2z})=\{(0.3,0.3),(0.3,-0.3),(-0.3,-0.3),(0.8,0.8),
(0.8,-0.8),(-0.8,-0.8)\}$. The faithfulness of a pair of waveforms
$(h_1,h_2)$ is quantified by the noise-weighted version of the overlap integral
\ref{eq:overlap}, which we rewrite in terms of $\epsilon$, the {\it mismatch} 
parameter (see e.g. \cite{Owen:1995tm})
\be
\label{eq:overlapnoise}
1-\epsilon\equiv\frac{\ds \int_0^\infty \frac{h_1(f) h_2^*(f)+h_1^*(f)h_2(f)}{S_n(f)} df}
{\ds 2\pa{\int \frac{|h_1(f)|^2}{S_n(f)} df}^{1/2}
\pa{\int \frac{|h_2(f)|^2}{S_n(f)} df}^{1/2}}\,,
\ee
where maximization over initial phase and arrival time is understood and 
$S_n(f)$ is the single-sided power spectral density of Advanced LIGO strain 
noise and waveforms are normalized with respect to noise weighted integrals.
The results of the above integrals comparing PhenSpin waveforms and long test 
waveforms with identical physical parameters are described in the next section.

\section{Results}
\label{se:result}

The analytical waveforms have been calibrated by comparison with 28 short test 
waveforms described in the previous section and the results obtained
by maximizing the overlap given by eq.~(\ref{eq:overlap}) are reported in 
tab.~\ref{tab:short}.
The determination of the $\omm$'s giving the best overlap for different spin 
values allowed to evaluate some of the coefficients in the phenomenological 
formulae (\ref{eq:omegamatch}) and (\ref{eq:fracmatch}), as given in 
tab.~\ref{tab:fit}.

\begin{figure}
\begin{center}
    \includegraphics[angle=90,width=.9\linewidth]{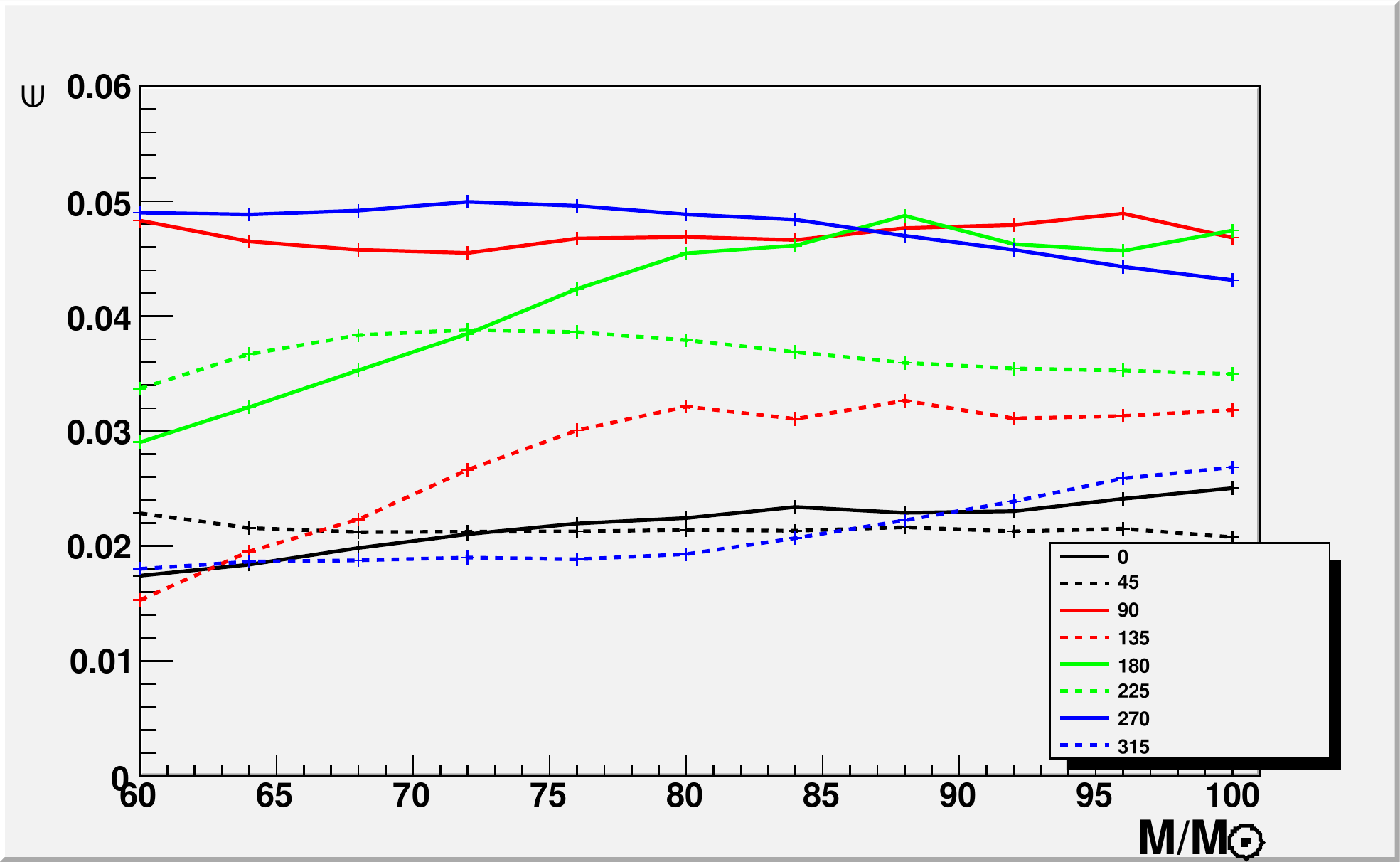}

    \includegraphics[angle=90,width=.9\linewidth]{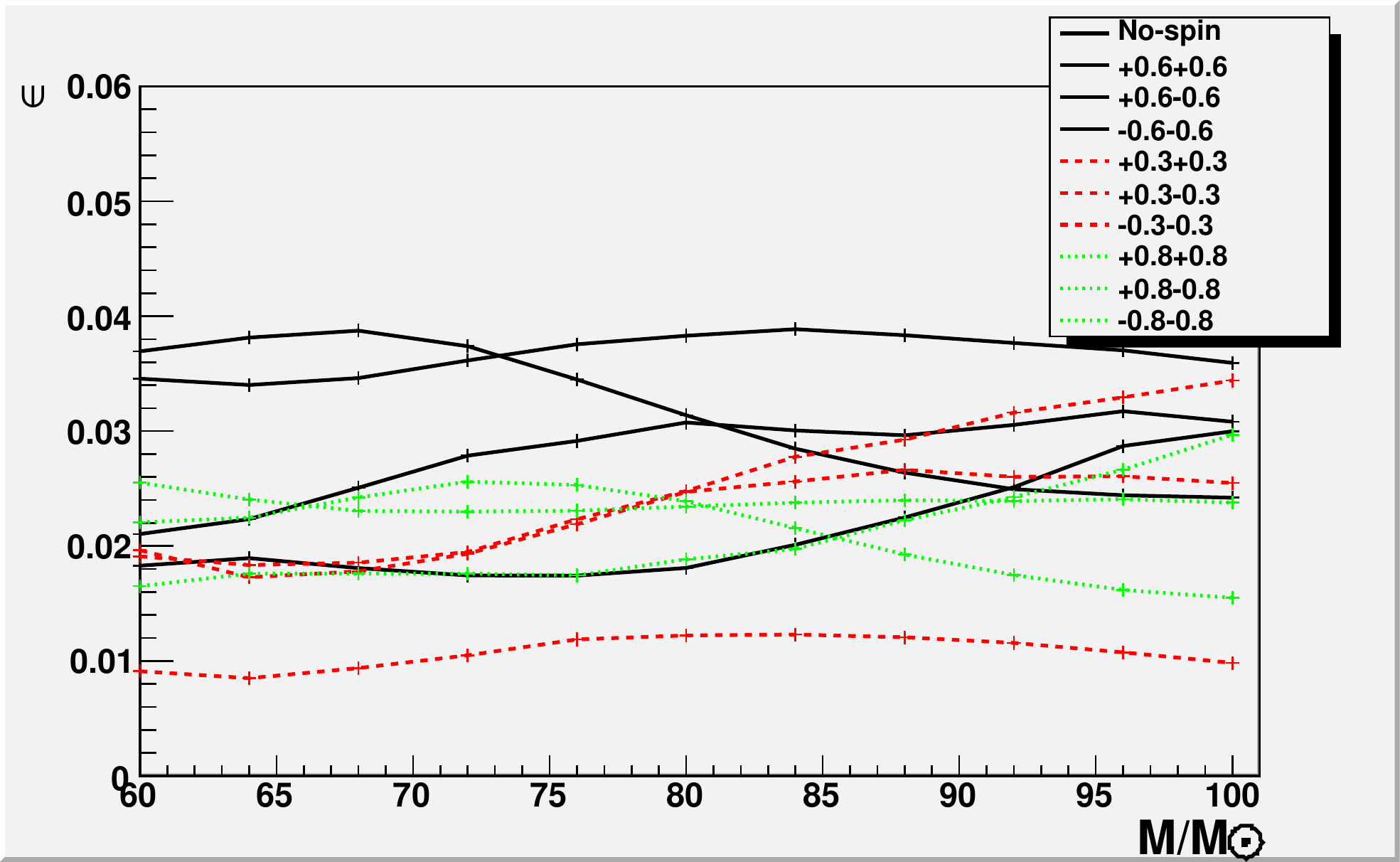}
    \caption{Quantitative comparison between the calibrated PhenSpin waveforms 
      and the numerical relativity ones (top) and phenomenological waveforms
      (bottom). The labels in the upper figure denote the angle (degrees) 
      between the initial $\mathbf S_1$ and the z-axis, the labels in the lower
      one denote the spin value along the z-axis of $\mathbf S_1$ and 
      $\mathbf S_2$ in units of $m_i^2$. }
    \label{fig:long}
\end{center}
\end{figure}

Once fixed the value of $\omm$ and $z$ for generic spin values, it is possible 
to generate analytical waveforms with any specific initial condition without 
any further tuning: the value of $\omm,z$ will be determined 
analytically via eqs.~(\ref{eq:omegamatch}) and (\ref{eq:fracmatch}) with the 
unknown coefficients arbitrarily set to zero.
It is then possible to generate waveforms with no tunable parameters for 
comparison with the second set of long waveforms.
The results of the faithfulness integrals described in 
eq.~(\ref{eq:overlapnoise}) are 
reported in fig.~\ref{fig:long} for the range of masses 60-100 $M_\odot$.
The lower range corresponds to the minimal mass value enabling the long 
waveforms to start at a physical frequency which is smaller than the lower
edge of the sensitive band (which we assume to be around 20Hz) whereas beyond 
the upper range value only the ring-down phase is in-band 
(for reasonable distances of the sources).

\section{Conclusions}
\label{se:concl}
We presented an analytical method to produce 
complete gravitational waveforms from spinning coalescing binaries.
The free parameters of the model are the values of the orbital
frequency at the transition from the inspiral to the phenomenological phase
and at the transition from the phenomenological to 
the ring-down phase.
After a calibration process involving the dominant multipolar mode obtained
by numerical relativity and other 
phenomenological construction from a different family, 
all the parameters have been
fixed and the PhenSpin are ready to use, once fed with the physical 
parameters of the coalescing binaries (mass, spins, inclination angles, initial
phase).
We computed noise-weighted overlap integral obtaining values between 0.95 and
0.98 for a wide range of masses.
Further investigations are necessary in order to assess how such mismatch can 
affect the precision of parameter estimation performed via these family
of waveforms.

\begin{center}
  \begin{table}
    \begin{tabular}{|c|c|c|c|c|}
      \hline
       \# &Overlap & $M\omm\times 10^{2}$ & $z$ & $M\omrd\times 10^{2}$\\      
      \hline
      1 & 0.992 & 5.50 & 0.76 & 29.6 \\
      \hline
      2 & 0.990 & 5.50 & 0.80 & 29.2 \\
      \hline
      3 & 0.989 & 5.50 & 0.85 & 28.8 \\
      \hline
      4 & 0.979 & 5.40 & 0.85 & 28.4 \\
      \hline
      5 & 0.976 & 5.75 & 0.85 & 27.9 \\
      \hline
      6 & 0.978 & 5.90 & 0.78 & 27.4 \\
      \hline
      7 & 0.986 & 5.95 & 0.85 & 26.9 \\
      \hline
      8 & 0.991 & 6.20 & 0.85 & 26.4 \\
      \hline
      9 & 0.992 & 6.15 & 0.84 & 26.0 \\
      \hline
      10 & 0.991 & 6.25 & 0.80 & 25.7 \\
      \hline
      11 & 0.993 & 6.20 & 0.85 & 25.6 \\
      \hline
      12 & 0.993 & 6.20 & 0.83 & 25.7 \\
      \hline
      13 & 0.993 & 6.15 & 0.81 & 26.0 \\
      \hline
      14 & 0.990 & 6.25 & 0.80 & 26.5 \\
      \hline
      15 & 0.983 & 6.20 & 0.80 & 27.2 \\
      \hline
      16 & 0.989 & 5.95 & 0.85 & 27.9 \\
      \hline
      17 & 0.985 & 6.15 & 0.80 & 28.6 \\
      \hline
      18 & 0.988 & 5.82 & 0.83 & 29.1 \\
      \hline
      19 & 0.984 & 6.00 & 0.78 & 29.6 \\
      \hline
      20 & 0.985 & 5.75 & 0.80 & 29.9 \\
      \hline
      21 & 0.984 & 5.50 & 0.83 & 30.1 \\
      \hline
      22 & 0.991 & 5.50 & 0.80 & 30.1 \\
      \hline
      23 & 0.992 & 5.45 & 0.77 & 30.1 \\
      \hline
      24 & 0.992 & 5.47 & 0.81 & 29.9 \\
      \hline
      25 & 0.992 & 5.55 & 0.84 & 26.9 \\
      \hline
      26 & 0.992 & 6.05 & 0.74 & 29.0 \\
      \hline
      27 & 0.993 & 5.20 & 0.85 & 26.9 \\
      \hline
      28 & 0.988 & 5.30 & 0.80 & 23.6 \\
      \hline
    \end{tabular}
    \caption{Values of the overlap integral given by eq.(\ref{eq:overlap}) 
      between the 
      analytical and the 28 test waveforms: the first 24 are numerically 
      generated, with initial conditions 
      ${\mathbf S_1}/m_1^2=(sin\,\alpha,0,cos\,\alpha)$
      and ${\mathbf S_2}/m_2^2=(-0.6,0,0)$, with 
      $\alpha= (k-1)\times \pi/12$ for 
      $0 < k \leq 24$, the following 4 have spins parallel to the orbital
      angular momentum with components respectively 
      $(S_{1z},S_{2z})/m^2=(0,0),(0.6,0.6),(0.6,-0.6),(-0.6,-0.6)$.
      In the columns from third to fifth the values of 
      $\omm$ and $z$ maximizing the overlap and of $\omrd$ are reported.}
    \label{tab:short}
  \end{table}
\end{center}

\begin{center}
  \begin{table}
    \begin{tabular}{|c|c|c|c|}
      \hline
      $\omm$-Coeff. & $\omm$ fit ($\times 10^{-3}$) & $z$-Coeff. & $z$ fit ($\times 10^{-2}$)\\
      \hline
      $a_0$ & 55.500 & $b_0$ & 84.00\\
      \hline
      $a_1$ & 0.997 & $b_1$ & -2.15\\
      \hline
      $a_3$ & -2.032 & $b_3$ & -4.42\\
      \hline
      $a_4$ & 5.629 & $b_4$ & -2.64\\
      \hline
      $a_5$ & 8.646 & $b_5$ & -5.88\\
      \hline
      $a_6$ & -5.909 & $b_6$ & -2.21\\
      \hline
      $a_7$ & 1.801 & &\\
      \hline
      $a_7$ & -14.059 & &\\
      \hline
      $a_8$ & 15.483 & &\\
      \hline
      $a_9$ & 8.922 & &\\
      \hline
    \end{tabular}
    \caption{Coefficients of eqs.~(\ref{eq:omegamatch}) and 
      (\ref{eq:fracmatch}) as determined by comparison of the analytical 
      waveforms with the 28 short test waveforms.}
    \label{tab:fit}
  \end{table}
\end{center}

\section*{Acknowledgments}
It is a pleasure to thank the organizers of the NRDA conference in Waterloo
for the stimulating scientific environment created by the meeting.  
This work was supported by the FAI grant from INFN and by the NSF grant 
PHY-0653550, PHY0955773, PHY-0925345, PHY-0941417 and PHY-0903973, PHY0955825, 
TG-PHY060013N.


\section*{References}
\begin{thebibliography}{99}

\bibitem{Abbott:2007kv}
  B.~Abbott {\it et al.} [ LIGO Scientific Collaboration ],
  Rept.\ Prog.\ Phys.\  {\bf 72 } (2009)  076901.
  [arXiv:0711.3041 [gr-qc]].

\bibitem{Acernese:2008zzf}
  F.~Acernese, M.~Alshourbagy, P.~Amico, F.~Antonucci, S.~Aoudia, K.~G.~Arun, P.~Astone, S.~Avino {\it et al.},
  Class.\ Quant.\ Grav.\  {\bf 25 } (2008)  184001.

\bibitem{Willke:2007zz}
  B.~Willke [ LIGO Scientific Collaboration ],
  Class.\ Quant.\ Grav.\  {\bf 24 } (2007)  S389-S397.

\bibitem{Blanchet:2006zz}
  L.~Blanchet,
  Living Rev.\ Rel.\  {\bf 9} (2006) 4.

\bibitem{Teukolsky:1973ha}
  S.~A.~Teukolsky,
  Astrophys.\ J.\  {\bf 185} (1973) 635.

\bibitem{Pretorius:2005gq}
  F.~Pretorius,
  Phys. \ Rev. \ Lett. {\bf 95} (2005) 121101 [arXiv: gr-qc/0507014].
  
\bibitem{Campanelli:2005dd}
  M.~Campanelli, C.O.~Lousto, P.~Marronetti, and Y.~Zlochower, 
  Phys. \ Rev. \ Lett.  {\bf 96} (2006) 111101 [arXiv:gr-qc/0511048].

\bibitem{Baker:2005vv}
  J.G.~Baker, J.~Centrella, D.~Choi, M.~Koppitz and J.~van Meter,
  Phys. \ Rev. \  Lett. {\bf 96} (2006) 111102 [arXiv:gr-qc/0511103].

\bibitem{Herrmann:2007ex}
  F.~Herrmann, I.~Hinder, D.M.~Shoemaker, P.~Laguna and R.A.~Matzner,
  Phys. \ Rev. \ D {\bf 76} (2007) 084032 [[arXiv:0706.2541]].

\bibitem{Tichy:2008du}
  W.~Tichy and P.~Marronetti,
  Phys.\ Rev.\ D {\bf 78} (2008) 081501 [arXiv:0807.2985 [gr-qc]].

\bibitem{Scheel:2008rj}
  M.A.~Scheel, M.~Boyle, T.~Chu, L.E.~Kidder, K.D.~Matthews, H.P.~Pfeiffer
  Phys.\ Rev.\ D {\bf 79} (2009) 024003 [arXiv:0810.1767 [gr-qc]].

\bibitem{Baker:2008mj}
  J.~G.~Baker, W.~D.~Boggs, J.~Centrella, B.~J.~Kelly, S.T.~McWilliams and J.R.~van Meter,
  Phys. \ Rev.\ D {\bf 78} (2008) 044046  [arXiv:0805.1428 [gr-qc]].

\bibitem{Ajith:2007qp}
  P.~Ajith {\it et al.},
  Class.\ Quant.\ Grav.\  {\bf 24} (2007) S689
  [arXiv:0704.3764 [gr-qc]];
  P.~Ajith {\it et al.},
  Phys. Rev. Lett. 106, 241101 (2011);
  arXiv:0909.2867 [gr-qc].

\bibitem{Lousto:2009ka}
  C.O.~Lousto, H.~Nakano, Y.~ Zlochower and M.~Campanelli,
  Phys.\ Rev.\  D {\bf 81} (2010) 084023 [arXiv:0910.3197 [gr-qc]].

\bibitem{Pollney:2007ss}
  D.~Pollney {\it et al.},
  Phys.\ Rev.\  D {\bf 76} (2007) 124002
  [arXiv:0707.2559 [gr-qc]].

\bibitem{Hannam:2009rd}
  M.~Hannam,
  Class.\ Quant.\ Grav.\  {\bf 26} (2009) 114001
  [arXiv:0901.2931 [gr-qc]].

\bibitem{Husa:2007zz}
  S. Husa, 
  Eur. Phys. J. ST {\bf 152} (2007) 183--207.

\bibitem{Pretorius:2007nq}
  F.~ Pretorius,
  ``Physics of Relativistic Objects in Compact Binaries: From Birth to 
  Coalescence'',
  Springer, Heidelberg (Germany), 2009 [arXiv:0710.1338].

\bibitem{Abadie:2011kd}
  J.~Abadie {\it et al.}  [The LIGO Scientific Collaboration and the Virgo
                  Collaboration and the Virgo],
  Phys.\ Rev.\  D {\bf 83} (2011) 122005
  [arXiv:1102.3781 [gr-qc]].

\bibitem{Buonanno:1998gg}
  A.~Buonanno and T.~Damour,
  Phys.\ Rev.\  D {\bf 59} (1999) 084006
  [arXiv:gr-qc/9811091];
  A.~Buonanno, T.~Damour,
  Phys.\ Rev.\  {\bf D62}, 064015 (2000)
  [arXiv: gr-qc/0001013];
  T.~Damour, P.~Jaranowski, G.~Schaefer,
  Phys.\ Rev.\  {\bf D62}, 084011 (2000)
  [arXiv: gr-qc/0005034].

\bibitem{Buonanno:2006ui}
  A.~Buonanno, G.~B.~Cook and F.~Pretorius,
  Phys.\ Rev.\  D {\bf 75} (2007) 124018
  [arXiv:gr-qc/0610122].

\bibitem{Buonanno:2007pf}
  A.~Buonanno, Y.~Pan, J.~G.~Baker, J.~Centrella, B.~J.~Kelly, S.~T.~McWilliams and J.~R.~van Meter,
  Phys.\ Rev.\  D {\bf 76} (2007) 104049
  [arXiv:0706.3732 [gr-qc]].

\bibitem{Damour:2009kr}
  T.~Damour and A.~Nagar,
  Phys.\ Rev.\  D {\bf 79} (2009) 081503
  [arXiv:0902.0136 [gr-qc]].

\bibitem{Pan:2009wj}
  Y.~Pan, A.~Buonanno, L.~T.~Buchman, T.~Chu, L.~E.~Kidder, H.~P.~Pfeiffer and M.~A.~Scheel,
  arXiv:0912.3466 [gr-qc].

\bibitem{Sturani:2010yv}
  R.~Sturani, S.~Fischetti, L.~Cadonati {\it et al.},
  J.\ Phys.\ Conf.\ Ser.\  {\bf 243}, 012007 (2010).
  [arXiv:1005.0551 [gr-qc]].

\bibitem{Pan:2003qt}
  Y.~Pan, A.~Buonanno, Y.~-b.~Chen {\it et al.},
  Phys.\ Rev.\  {\bf D69 } (2004)  104017;
  [gr-qc/0310034].

\bibitem{Buonanno:2004yd}
  A.~Buonanno, Y.~-b.~Chen, Y.~Pan, M.~Vallisneri,
  Phys.\ Rev.\  {\bf D70 } (2004)  104003;
  [gr-qc/0405090].

\bibitem{Brown:2007se}
  D.~A.~Brown, J.~Crowder, C.~Cutler {\it et al.},
  Class.\ Quant.\ Grav.\  {\bf 24 } (2007)  S595-S606.
  [arXiv:0704.2447 [gr-qc]].

\bibitem{lal}
  https://www.lsc-group.phys.uwm.edu/daswg/projects/lal.html

\bibitem{Buonanno:2009zt}
  A.~Buonanno, B.~Iyer, E.~Ochsner, Y.~Pan and B.~S.~Sathyaprakash,
  Phys.\ Rev.\  D {\bf 80} (2009) 084043
  [arXiv:0907.0700 [gr-qc]].

\bibitem{Arun:2008kb}
  K.~G.~Arun, A.~Buonanno, G.~Faye and E.~Ochsner,
  Phys.\ Rev.\  D {\bf 79} (2009) 104023
  [arXiv:0810.5336 [gr-qc]].

\bibitem{Mikoczi:2005dn}
  B.~Mikoczi, M.~Vasuth, L.~A.~Gergely,
  Phys.\ Rev.\  {\bf D71 } (2005)  124043.
  [astro-ph/0504538].

\bibitem{Racine:2008kj}
  E.~Racine, A.~Buonanno and L.~E.~Kidder,
  Phys.\ Rev.\  D {\bf 80} (2009) 044010
  [arXiv:0812.4413 [gr-qc]].

\bibitem{Blanchet:2011zv}
  L.~Blanchet, A.~Buonanno and G.~Faye,
  arXiv:1104.5659 [gr-qc].

\bibitem{Baker:2006ha}
  J.~G.~Baker, J.~R.~van Meter, S.~T.~McWilliams, J.~Centrella and B.~J.~Kelly,
  Phys.\ Rev.\ Lett.\  {\bf 99} (2007) 181101
  [arXiv:gr-qc/0612024].

\bibitem{Boyle:2007ft}
  M.~Boyle {\it et al.},
  Phys.\ Rev.\  D {\bf 76} (2007) 124038
  [arXiv:0710.0158 [gr-qc]].

\bibitem{Berti:2005ys}
  E.~Berti, V.~Cardoso and C.~M.~Will,
  Phys.\ Rev.\  D {\bf 73} (2006) 064030
  [arXiv:gr-qc/0512160].

\bibitem{Barausse:2009uz}
  E.~Barausse and L.~Rezzolla,
  Astrophys.\ J.\  {\bf 704} (2009) L40
  [arXiv:0904.2577 [gr-qc]].
 
\bibitem{Schnetter-etal-03b}
  E.~Schnetter,  S.H.~Hawley and I.~Hawke,
  Class. Quant. Grav  (2004) {\bf  21} 1465--1488 [arXiv:0310042 [gr-qc]].

\bibitem{cactus-web}
  Cactus Computational Toolkit home page: {\tt http://www.cactuscode.org}.

\bibitem{Husa:2004ip}
  S.~Husa, I.~Hinder and C.~Lechner,
  Computer Physics Communications {\bf 174} (2006) 983-1004 [arXiv:0404023 [gr-qc]].

\bibitem{Fischetti:2010hx}
  S.~Fischetti, J.~Healy, L.~Cadonati, L.~London, S.~R.~P.~Mohapatra, D.~Shoemaker,
  Phys.\ Rev.\  {\bf D83 } (2011)  044019.
  [arXiv:1010.5200 [gr-qc]].

\bibitem{Leaver:1985ax}
  E.~W.~Leaver,
  Proc.\ Roy.\ Soc.\ Lond.\  {\bf A402 } (1985)  285-298;
  E.~Berti, V.~Cardoso, C.~M.~Will,
  Phys.\ Rev.\  {\bf D73 } (2006)  064030.
  [gr-qc/0512160].

\bibitem{Owen:1995tm}
  B.~J.~Owen,
  Phys.\ Rev.\  {\bf D53}, 6749-6761 (1996).
  [gr-qc/9511032].
  
\end{thebibliography}
\end{document}